\documentclass[aps,prd,showpacs,preprintnumbers,nofootinbib,floatfix,floats,groupedaddress]{revtex4}
\usepackage{bm}
\usepackage{latexsym}
\usepackage{tabularx}
\usepackage{dcolumn}
\usepackage{amsmath,amsfonts,amssymb}
\usepackage{graphicx,epsfig}
\usepackage{color}
\usepackage{subfigure}
\usepackage{sidecap}
\usepackage{longtable}
\newcommand{\de}{{\rm d}}
\newcommand{\bea}{\begin{eqnarray}}
\newcommand{\eea}{\end{eqnarray}}
\newcommand{\f}{\frac}
\newcommand{\T}{\rule{0pt}{3.6ex}}

\newcommand{\HS}{\rule{8.5pt}{0ex}}
\newcommand{\HSN}{\rule{4.5pt}{0ex}}
\newcommand{\B}{\rule[-1.0ex]{0pt}{0pt}}

\begin{document}

\title{Weighing neutrinos using high redshift galaxy luminosity functions}

%##############################################################################################
\author{Charles Jose$^1$}
\email{charles@iucaa.ernet.in}
\author{Saumyadip Samui$^2$}
\email{ssamui@gmail.com}
\author{Kandaswamy Subramanian$^1$}
\email{kandu@iucaa.ernet.in}
\author{Raghunathan Srianand$^1$}
\email{anand@iucaa.ernet.in}
\affiliation{1. IUCAA, Pune University Campus, Ganeshkhind,
 Pune 411007, INDIA.}
\affiliation{2. SISSA, via Bonomea, 265, 34136 Trieste, Italy}
%#############################################################################################

\date{\today}

%#############################################################################################
\clearpage

\begin{abstract}{
Laboratory experiments measuring neutrino oscillations, 
indicate small mass differences between different mass eigenstates of neutrinos.
The absolute mass scale is however not determined, with at present
the strongest upper limits coming from astronomical observations rather
than terrestrial experiments.
The presence of massive neutrinos suppresses the growth
of perturbations below a characteristic mass scale, thereby 
leading to a decreased abundance of collapsed dark matter halos.
Here we show that this effect can significantly alter the predicted 
luminosity function (LF) of high redshift galaxies. In particular
we demonstrate that a stringent constraint on the neutrino mass 
can be obtained using the well measured galaxy LF and
our semi-analytic structure formation models.  
Combining the constraints from the 
Wilkinson Microwave Anisotropy Probe 7 year (WMAP7) data
with the  LF data at $z \sim 4$, we get a limit
on the sum of the masses of 3 degenerate neutrinos
$\Sigma m_\nu < 0.52$ eV at the 95 \% CL. 
The additional constraints using the prior on 
Hubble constant 
strengthens this limit to $\Sigma m_\nu \le 0.29$ eV
at the 95 \% CL.
This neutrino mass limit is a factor $\sim4$ 
improvement compared to the constraint based on the 
WMAP7 data alone, and as stringent as known
limits based on other astronomical observations.
As different astronomical measurements may suffer from
different set of biases, the method presented here
provides a complementary probe of $\Sigma m_\nu$. 
We suggest that repeating this exercise 
on well measured luminosity functions over 
different redshift ranges can provide independent and 
tighter constraints on $\Sigma m_\nu$.}  
\end{abstract}

%\arxivnumber{}

\clearpage

\maketitle
%#############################################################################################

%##############################################################################################

\section{Introduction}

\par A cosmic background of neutrinos is one of the key 
predictions of standard cosmology.
Their predicted abundance is comparable to that of the relic photons. 
Thus if neutrinos have a mass they can contribute 
significantly to the matter density in the universe. 
Experiments which detect neutrino oscillations
have measured small but non-zero differences between the mass eigenstates
of neutrinos, with at least one of them having a mass larger than
$\sim 0.05$ eV \citep{maltoni_2004,GMS10}.
The absolute mass scale of 
neutrinos could be inferred from various $\beta$-decay experiments 
\citep{ott_wei_2008,vogel_2008}. However at present stronger constraints 
on neutrino mass are obtained 
from cosmological observations. In particular observations related to 
anisotropies in the 
Cosmic Microwave Background Radiation (CMBR) and the growth of structure in the universe
can play an important role (see \cite{les_pas_2006,hannestad_2010} for reviews). 
For example the recent 
WMAP7 data itself has been used to set an 
upper limit on the of sum of neutrino masses, 
$\Sigma m_{\nu}< 1.15-1.3$ eV,
for sudden reionization \cite{kom_smi_2010,arch10} and  
$\Sigma m_{\nu}< 1.7$ eV for a generalized reionization scenario
\cite{arch10}.
Thus one is
dealing with the universe having matter density dominated by
cold dark matter (CDM) with massive neutrino (Hot dark matter, HDM)
providing sub-dominant contribution
(i.e the neutrino density parameter
$\Omega_\nu = \Sigma m_\nu/93.14 h^2 {\rm eV} \le 0.014h^{-2}$, where
$h$ is the Hubble constant H$_0$ in units of $100$ km s$^{-1}$ Mpc$^{-1}$;
see chapter 2 of Ref.~\cite{dodelson_2003}).
This is usually referred to as the mixed dark matter (MDM) scenario.

Interestingly, the presence of even such a subdominant 
HDM component leads to the suppression of the growth of density perturbations 
below a scale known as the free streaming scale \cite{bon_sza_1983}. 
This suppression occurs because neutrinos escape
(or free-stream out of) their own density
perturbations below the free-streaming scale. Thus only the
CDM component can lead to perturbation growth
below this scale. This free streaming scale is time dependent
and also depends on the neutrino mass.
The suppression
of the growth of density perturbations below the free
streaming scale leads to a reduction
in the matter power spectrum
and a delay in the
formation of structures in the universe.
This in turn results in a reduced abundance of
dark matter halos at any given epoch, 
above a characteristic mass scale.
Thus observations related to large scale structure formation
in the universe can be used to probe the absolute mass scale of
neutrinos.
It has been shown that sub eV constraints are obtained
on neutrino mass by combining CMB data with currently available data on galaxy 
surveys at low redshift  \cite{tho_abd_2010},
counts of low redshift massive galaxy clusters \cite{man_all_2010}, 
inter-galactic medium (IGM)
Lyman-$\alpha$ absorption power spectrum \cite{seljak_2006,veil_2010}
and weak lensing \cite{ter_sch_2008}.
Unlike CMB observations all the astrophysical observations are
affected by different systematics related to baryonic physics.
Thus it is important to explore whole range of observables
(with different biases) to get realistic constraints on $\Sigma m_\nu$.
There is a growing wealth of observations on 
high redshift galaxies which may also provide independent and equally 
useful constraints on neutrino mass. 
Thus in this work we explore the possibility of using the luminosity functions (LF)
of high redshift Lyman break galaxies (LBGs) for constraining $\Sigma m_\nu$.

The basic idea is as follows: The reduction in the matter power spectrum
in models with $\Sigma m_\nu > 0$, compared to models where $\Sigma m_\nu=0$
implies a reduced abundance of galactic scale dark matter halos
at high redshifts.
In order to account for the observed LF of
these galaxies (in number per unit volume per unit luminosity),
the light to mass ratio of each galactic halo has to
be systematically higher in the models with $\Sigma m_\nu > 0$.
However changing the light to mass ratio is degenerate with
the unknown extinction correction one applies to the
observed LF.
Nevertheless, this degeneracy can be lifted if one has a
feature in the LF at some characteristic mass scale,
introduced by various feedback processes like the
radiative feedback after reionization.
In such cases
the shape of the predicted galaxy luminosity
function depends on the neutrino mass.
We use this idea to constrain neutrino mass.
In the next section we discuss structure formation 
models incorporating massive neutrinos.
Section~\ref{section:LF} describes  
our semi-analytic models for the UV luminosity functions. The effect
of a non-zero neutrino mass on the high redshift LFs is
studied in Section~\ref{nuLF}. We present
our limits on $\Sigma m_\nu$ in Section~\ref{result} using a
Markov Chain Monte Carlo analysis and
conclude in Section~\ref{conclu}.

%%%%%%%%%%%%%%%%%%%%%%%%%%%%%%%%%%%%%%%%%%%%%%%%%%%%%%%%%%%%%%%%%%%%%%%%%%%%%%%%%%%%%%%%%%%%

\section{Structure formation in MDM cosmology}

%\subsection{Linear evolution of perturbations}

%##############################################################################################

A crucial ingredient of any model of structure formation is $\sigma(M,z)$,
the rms density fluctuations on any mass scale $M$ as a function of
redshift $z$. This is given by
\begin{equation}
\sigma^2(M,z) = \sigma^2(R,z)
= \int_0^\infty{\frac{dk}{2\pi^2} k^2  P_k(z) W^2(k,R)} \label{eqn:sigma}
\end{equation}
where $R$ is the comoving radius of a sphere containing mass $M$,
$k$ is the comoving wave number, $W(k,R)$ is the top hat window function
in Fourier space and $P_k(z)$ is the linear power spectrum of the
density fluctuations at $z$. For a universe with massive neutrinos
$P_k(z) = T^2(k,z,z_i;m_\nu) P_k(z_i)$.
Here $P_k(z_i)$ is the initial power spectrum at $z_i$ and
the function $T(k,z,z_i;m_\nu)$ is the matter transfer function
in a $\Lambda$CDM universe with
massive neutrinos.

\par 
The transfer function in mixed dark matter (MDM) 
cosmology with neutrinos 
has been studied extensively \cite{hu_eis_1998,eis_hu_1999,bon_sza_1983}. 
Eisenstein and Hu \cite{eis_hu_1999} give a fitting formula of the form 
\begin{equation}
T(k,z,z_i;m_\nu) = T_{master}(k;m_\nu)D(k,z,z_i;m_\nu). 
\label{eqn:transfer}
\end{equation}
The function $D(k,z,z_i;m_\nu)$ 
is the scale dependent growth 
factor of CDM, baryon and neutrino perturbations in a $\Lambda$CDM universe in presence 
of free streaming neutrinos and  $T_{master}$ is a master-transfer function. 
The explicit forms of both these functions are given by Eisenstein and Hu \cite{eis_hu_1999} 
(see also \cite{hu_eis_1998}).  
Their fit is optimized for a total number of three neutrinos which include massive as 
well as massless species. 
Note that the standard model of particle physics predicts the effective number of neutrinos 
($N_\nu$) to be 3.04  with the 0.04 coming from incomplete neutrino freeze-out and finite 
temperature effects around $e^+$ $e^-$ annihilation \cite{hannestad_2010}.
We have checked that this transfer function is in
good agreement with $T$ computed numerically
using the 'CAMB' code \cite{camb_2000} (Lewis, Challinor \& Lasenby 2000).
We will also fix $N_\nu=3$ and adopt this transfer function 
while calculating the luminosity functions below.

Given the transfer function and $\sigma(M,z)$ one can 
estimate the abundance of dark matter halos, for example
using the Press-Schechter (PS) \cite{pre_sch_1974} approach. 
This also requires one to specify $\delta_c$,
the linearly extrapolated critical density contrast 
needed for collapse of
a spherical top hat over dense region.
For a flat universe with only CDM, the critical density contrast of 
collapse is 1.686 \cite{peebles_1980,pad_kan_1992}. 
The calculation of $\delta_c$ for the 
$\Lambda$CDM model without any massive neutrino species, 
has been done before by \cite{eke_col_1996}, 
who found that $\delta_c$ for this model is very nearly the same 
as that for the pure CDM model. 
However, to our 
knowledge, a corresponding estimate of 
$\delta_c$ does not exist for models with massive neutrinos. 
As the abundance of halos is exponentially sensitive to the value of 
$\delta_c$ when using the PS formalism, it is important
to estimate it even in MDM models. We do this below using the
spherical model for nonlinear evolution adopting 
a flat $\Lambda CDM$ universe with massive neutrinos.

\subsection{Spherical model for nonlinear collapse in MDM cosmology}

The spherical model or top hat model \cite{gun_got_1972}, 
gives the nonlinear growth of a 
uniformly overdense spherical region in a smooth background of expanding 
universe. In spherical model 
we study the evolution of physical density contrast $\delta(r,z)$ in real 
space directly rather than the evolution of it's Fourier components 
$\delta(k,z)$. For this purpose, we assume a 
spherical uniformly over dense region in the background expanding universe. 
Below the free-streaming scale of the neutrinos, only the CDM and baryons 
can cluster due to gravity, although both the cosmological constant 
$\Lambda$ and neutrinos will contribute to the expansion. 
Note that the mass associated
with the free streaming scale is much larger 
than the galactic scales we consider
below. Thus the clustering mass is solely contributed by 
the baryons and the CDM.
This is also borne out by recent simulations of structure
formation in an MDM universe by Brandbyge et al \cite{bra_han_2010}. 

The gravitational acceleration of the shell of radius $r$ 
(with initial radius $r_i$) in the spherical model is then given by
\begin{equation}
\ddot{r} = -\frac{G M(z_i)}{r^2} - \frac{4}{3}\pi G (\rho_\nu(z) - 2\rho_{\Lambda})r .
\label{eqn:lcdmn_sphere}
\end{equation}
Here $M$ is the total mass which can cluster within the spherical region, 
and hence does not include
any contribution from the neutrinos. Explicitly we have
\begin{equation}
M = 4 \pi r^3_i (\rho_{CDM}(z_i)+ \rho_b(z_i)) (1+\delta_{cb}(z_i))/3 = 
4 \pi r^3_i (1-f_\nu) \rho_m(z_i)(1+\delta_{cb}(z_i))/3.
\end{equation}
Here $\rho_{CDM}(z)$, $\rho_b(z)$, $\rho_\nu(z)$ and $\rho_\Lambda(z)$ are respectively the
background densities of CDM, baryons, neutrinos and cosmological constant
at redshift $z$. We
have also defined the neutrino fraction 
$f_\nu= \rho_\nu/\rho_m$, with 
$\rho_m = \rho_{CDM} + \rho_b + \rho_\nu$ is the total matter density.
Further $\delta_{cb} = (\delta\rho_{CDM} + \delta\rho_{b})/(\rho_{CDM} + \rho_{b}) $
is fractional overdensity in the CDM $+$ baryon
component within the spherical region.
Note that the total density contrast $\delta_m = \delta \rho_m/ \rho_m 
= \delta_{cb} (1-f_\nu)$, since $\delta\rho_\nu=0$ inside the spherical region.
Even though neutrinos do not cluster below their free-streaming scale, 
their uniform density does lead to a deceleration of the shell, 
through the $\rho_\nu(z)$ term in Eq.~(\ref{eqn:lcdmn_sphere}).
Moreover the cosmological constant leads to a positive acceleration of the shell proportional
to $(\rho_\Lambda+3p_\Lambda)$, where the pressure $p_\Lambda = -\rho_\Lambda$. 
At any redshift $z$ and $\rho_\Lambda(z)=\rho_{c} \Omega_{\Lambda}(0)$ and
$\rho_\nu(z) = \rho_c \Omega_\nu (1+z)^3$, where $\Omega_i$ is the present density
of component `$i$' in units of the critical density $\rho_c$.

The Hubble expansion rate at any redshift is given by
\begin{equation}
H(z) = H(0)\left(\Omega_m (1+z)^3 + \Omega_{\Lambda} \right)^{1/2} \label{eqn:hubble}
\end{equation}
where $H(0)$ is the present value of Hubble constant. Noting that 
\begin{equation}
dt =\frac{}{} \frac{dt}{da} \frac{da}{dz} dz 
= -\frac{dz}{H(z) (1+z)^3}, \label{eqn:dt}
\end{equation}
we can numerically solve Eqn. (\ref{eqn:lcdmn_sphere}) 
for a given set of initial 
conditions to obtain trajectory of the shell $r(z)$, and the 
redshift of collapse $z_c$.
We assume that initially the density contrast 
$\delta_{cb}(z_i)=\delta_i$ is small 
enough that the overdense region is expanding along with the background. 
Thus the initial velocity of the shell at radius $r$ is 
$v_i=H(z_i) r(z_i)$.

\par
Once we know the redshift of collapse $z_c$, we can evolve linear equations of 
perturbations using the same initial conditions to calculate the linearly 
extrapolated critical 
density contrast, $\delta_c$ at $z_c$. This linear evolution of $\delta_{cb}$ 
for scales much below the free streaming scale is governed by
\begin{equation}
\ddot\delta_{cb} + 2 H \dot\delta_{cb} - 4 \pi G (1-f_\nu) \Omega_m \rho_c \delta_{cb} = 0,
\label{linevol} 
\end{equation}
whose solution can only be determined 
numerically for a $\Lambda$+MDM universe. 

\begin{figure}
\begin{center}
{\includegraphics[trim=0cm 0cm 0cm 0cm, clip=true, width =9cm, height=9.5 cm, angle=0] 
{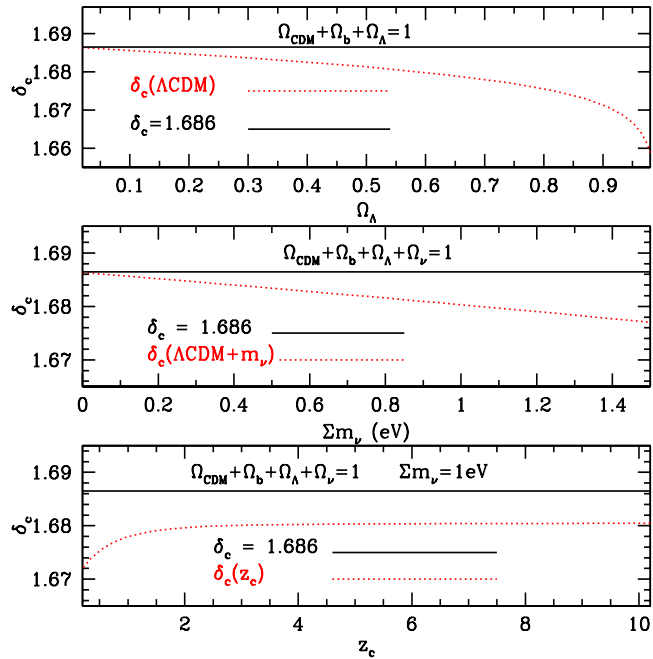}}
\caption{Dependence of $\delta_c$ on various parameters. 
The top panel shows $\delta_c$ as a function of $\Omega_\Lambda$ 
for a $\Lambda$CDM universe without 
any massive neutrinos, with $z_c=0$. The middle panel shows 
$\delta_c$ as a function of $\Sigma m_\nu$ 
for $\Lambda$CDM universe with massive neutrinos where $z_c=4$. 
The cosmological parameters 
are chosen according to our fiducial model ($\Omega_\Lambda=0.735$, $h=0.71$) 
and by keeping $\Omega_{CDM} + \Omega_b + \Omega_\nu = 0.265$.
In the bottom panel the dependence of $\delta_c$ on $z_c$
is plotted for $\Sigma m_\nu=1$ eV and for the cosmological
parameters as above.
In all the panels the solid horizontal (solid dark) line at 1.686 
corresponds to $\delta_c$ for a universe with only CDM in it. 
The red dotted line shows the variation in $\delta_c$ with the 
parameter of interest.
}
\label{fig.spherical_collapse}
\end{center}
\end{figure}

In the upper panel of Fig.~\ref{fig.spherical_collapse}, we give
$\delta_c$ for $z_c=0$ as a function of $\Omega_\Lambda$, 
as determined by the above procedure, assuming zero mass for the neutrinos.
We confirm the weak dependence of $\delta_c$ 
on $\Omega_{\Lambda}$ (with $\delta_c$ differing from
$1.686$ only by $0.5\%$ for $\Omega_\Lambda=0.7$), 
as also found by \cite{eke_col_1996}.
Note that even smaller changes in $\delta_c$ ($\sim 0.006\%$) 
would obtain in a $\Lambda$CDM model for the 
higher collapse redshifts 
(say $z_c =4$) relevant for our work.
In the middle panel of  Fig.~\ref{fig.spherical_collapse}, 
we show the $\delta_c$ dependence on $\Sigma m_\nu$, 
for $z_c=4$. We have adopted the cosmological parameters
of a fiducial model with $\Omega_\Lambda=0.735$ and $h=0.71$. 
The change in $\delta_c$ 
is very small even in the case of universe with massive neutrinos
for the $\Sigma m_\nu$ range constrained by CMBR observations.
For example, with $z_c=4$, $\delta_c$ decreases only 
to $1.68$ for $\Sigma m_\nu=1$ eV, or only a decrease of $\sim 0.4\%$
from the canonical value. 
The bottom panel gives $\delta_c$ as a function of $z_c$ 
for $\Sigma m_\nu =1$ eV and for the cosmological
parameters as above.
At $z_c\sim 0$ now $\delta_c$ decreases to $1.671$ 
(or a fractional change of $\sim 0.9\%$),
due to the additional effect of
the $\Omega_\Lambda$ at such low redshifts.
Based on the above discussion, it is a reasonable approximation 
to take the fiducial value of $\delta_c=1.686$, when computing 
the abundance of halos at high redshifts, even in the
presence of massive neutrinos.

%##########################################################################################

\subsection[]{The abundance of halos}

%%%%%%%%%%%%%%%%%%%%%%%%%%%%%%%%%%%%%%%%%%%%%%%%%%%%%%%%%%%%%%%%%%%%%%%%%%%%%%%%%%%%%%%%%%%%
\begin{figure*}
\begin{center}
\includegraphics[trim=0cm 0cm 0cm 0cm, clip=true, width =16.5cm, height=12.5cm, angle=0]
{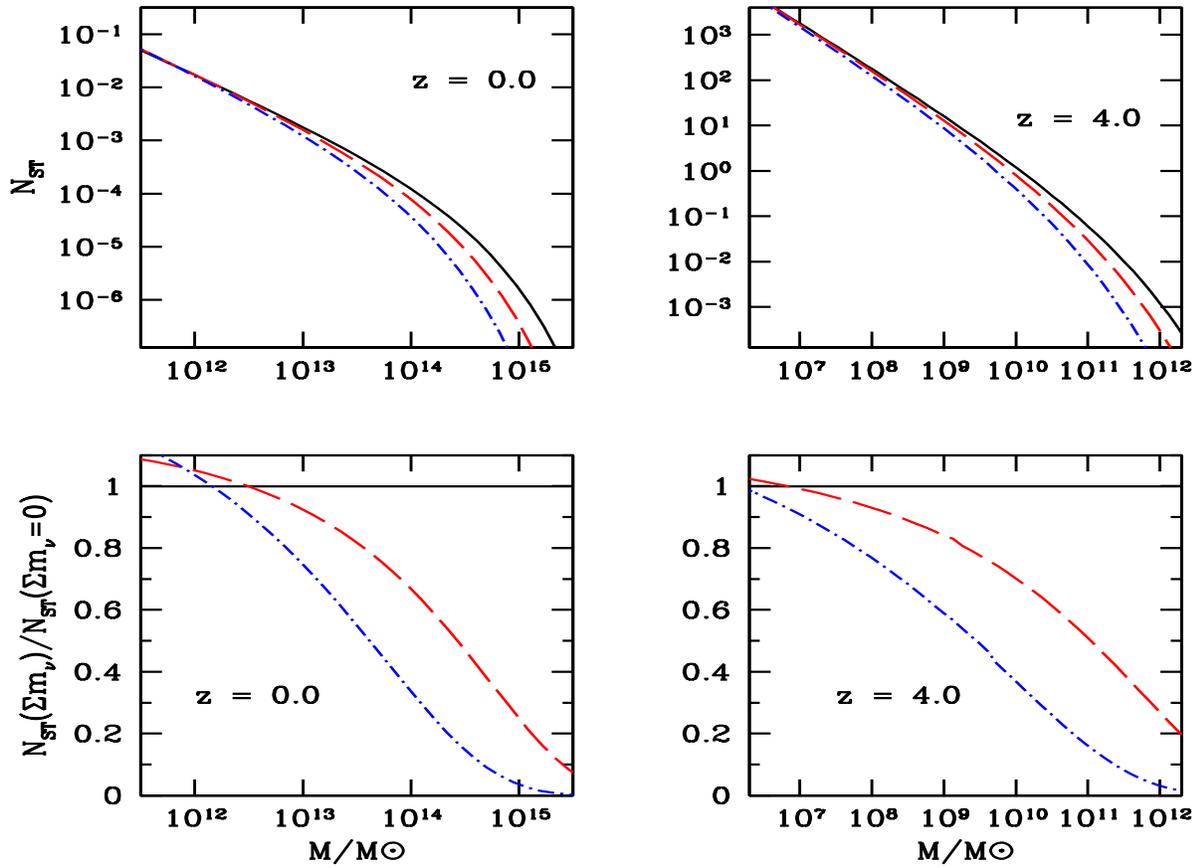}
\caption{The halo abundance $N_{ST}$ for $\Lambda$CDM and $\Lambda$MDM models. 
The top panels show $N_{ST}$ as a function of halo mass 
for $\Sigma m_\nu=0$ (solid black lines), $\Sigma m_\nu=0.5$ eV 
(long dashed red lines) and $\Sigma m_\nu=1$ eV (dash-dotted blue lines). 
The bottom panels show the ratio of $N_{ST}$ 
for different values of $\Sigma m_\nu$ to that obtained taking 
$\Sigma m_\nu=0$. The left panels
are for $z=0$, while the right panels are for $z=4$.
}\label{fig.Nst} 
\end{center}
\end{figure*}

The abundance of dark matter halos as a function of mass and 
redshift plays a crucial role 
in understanding various aspects of galaxy formation. 
This halo mass function can be obtained 
through analytical formulations as well as 
N-body simulations. The first 
analytical form of halo mass function was given by  
Press and Schecter \cite{pre_sch_1974}. 
Subsequently several alternative mass functions have been suggested which
better fit the results of N-body simulations, a popular choice being the Sheth-Tormen (ST)
mass function \cite{she_tor_1999}.
These mass functions assume the initial density field to be Gaussian 
random. And also that
a region of size $R$ (containing a mass $M$) collapses when its linearly 
extrapolated density contrast reaches a critical density $\delta_c$,  
as predicted by the spherical model of nonlinear evolution. 
We found above that $\delta_c$ for the CDM+baryon component 
is only slightly altered from the canonical value of $1.686$, 
by the presence of massive neutrinos. 

Brandbyge et al \cite{bra_han_2010} have shown that 
the ST mass function provides an excellent fit for mass functions 
of dark matter halos obtained from N-body simulations including massive
neutrinos. This has also been confirmed recently 
by Marulli et al \cite{marulli_2011}.
In the ST formula the comoving number density of halos with mass between M 
and M+dM at any redshift z is given by
\begin{equation}
N_{ST}(M,z)dM = \dfrac{\rho}{M} A \sqrt{\dfrac{2a}{\pi}} 
  \left[ 1 + \left(\dfrac{\sigma^2}{a\delta^2_c} \right)^p \right] 
  \dfrac{\delta_c}{\sigma} 
\left(\frac{-1}{\sigma} \frac{d\sigma}{dM} \right) 
\exp \left( \dfrac{-a \delta_c^2}{2\sigma^2} \right)
dM 
\label{eqn:ST_no}
\end{equation}
where from the numerical fitting 
$A=0.3222$, $a=0.707$, $p=0.3$ and $\delta_c=1.686$. 
Brandbyge et al \cite{bra_han_2010} also find that 
one needs to use $\rho=\rho_{cb}= (\rho_{CDM} + \rho_b)$
and calculate the mass as $M = M_{cb} = 4 \pi r^3 (\rho_{CDM} + \rho_b)/3$
in using ST formula to fit the halo abundance.
However they calculate $\sigma$ using the total matter power
spectrum \cite{bra_han_2010} (also Brandbyge; private communication),
although naively one may have expected that $\sigma$ should be calculated
using the power spectrum of CDM+baryons 
\footnote{
Note that, for length scales much smaller than the free-streaming scale, 
since $\delta_m = \delta_{cb}(1-f_\nu)$,
$\sigma$ calculated from the total matter power
spectrum is $\approx \sigma_{cb}(1-f_\nu)$, where $\sigma_{cb}$ 
is that calculated from the power spectrum of the CDM+baryons. 
}.
In our work we follow the prescription of \cite{bra_han_2010} in calculating
the halo abundance, as this seems to fit the simulated
data well.

We illustrate the effect of massive neutrinos on the abundance of dark matter
halos in Fig.~\ref{fig.Nst}, where we plot in the upper panels, 
$N_{ST}$, using the prescription 
of \cite{bra_han_2010}, for two redshifts $z=0$ and $z=4$.
The solid black line in Fig.~\ref{fig.Nst} gives $N_{ST}$, for 
a fiducial $\Lambda$CDM model with $\Sigma m_\nu=0$.
As the fiducial $\Lambda$CDM model, 
we adopt a set cosmological parameters consistent with WMAP-CMBR 7 year data 
\cite{wmap7}. Thus we adopt 
$h = 0.71$, $\Omega_b = 0.0448$ and $\Omega_{CDM} = 0.220$. 
The initial power spectrum is specified by the best 
fit values of scalar spectral 
index $n_s = 0. 953$ and curvature fluctuation 
amplitude $\Delta_R^2 = 2.43 \times 10^{-9}$ 
at a pivot scale $k_0 = 0.002 Mpc^{-1}$ (or a $\sigma_8 = 0.801$). 
We then examine models with non-zero $\Sigma m_\nu$ 
by varying $\Omega_{CDM}$ and satisfying the constraint that 
$\Omega_{CDM} + \Omega_\nu = 0.220$. 
The long dashed (red) line gives $N_{ST}$ 
for the case when $\Sigma m_\nu=0.5$ eV,
while the dash-dotted (blue) line is for the case $\Sigma m_\nu=1$ eV.
The bottom panels give the ratio of $N_{ST}$ between the non-zero
neutrino mass cases with respect to the fiducial model, as a function of $M$.

From this figure we can see that the number of collapsed halos 
decreases with increase in neutrino mass for 
halo mass scales above a characteristic mass scale $M_c$, 
while there is a slight increase in abundance of halos below this mass scale.
The characteristic mass $M_c \sim 10^{12}M_\odot$ at $z=0$ while 
$M_c \sim 1.6 \times 10^6 M_\odot$ at $z=4$ for $m_\nu =1$ eV.
The decrease in abundance of large mass halos arises 
as a result of the suppression of growth due to the presence of free-streaming neutrinos.  
In our model with $\Sigma m_\nu = 1.0$ eV, we find that at $z=0$, 
the abundance of galaxy clusters of mass
$M \sim 10^{14} M_\odot$ decreases by a factor $\sim 3.2$, while for bigger clusters with 
$M \sim 10^{15}M\odot$, the decrease in abundance is by a factor $\sim 25$.
Similarly at $z=4$, the abundance of galactic scale halos of mass 
$10^{11} M_\odot$ and $10^{12}M_\odot$ 
are suppressed respectively by factors $\sim 6.5$ and $\sim 30$, in models
with $\Sigma m_\nu = 1$eV compared to the zero neutrino mass case.
Thus strong constraints on the neutrino mass can be obtained from measuring 
the abundance of galaxy clusters at low redshifts or equally, 
the abundance of galaxies at high redshifts
as probed by their observed luminosity functions. 

%%%%%%%%%%%%%%%%%%%%%%%%%%%%%%%%%%%%%%%%%%%%%%%%%%%%%%%%%%%%%%%%%%%%%%%%%%%%%%%%%%%%%%%%%%%%%

\section{High redshift galaxy luminosity functions}
\label{section:LF}
In the previous section we have described how the presence of
massive neutrinos suppresses the high redshift galactic halo formation.
In this section, using semi-analytical models, we investigate 
the effect of massive 
neutrinos on high redshift galaxy UV LF. 
We model high redshift galaxy LF using the 
semi-analytical treatment of Samui, Srianand \& Subramanian (2007) 
\cite{sam_sri_2007} (hereafter SSS07) which were successful in explaining the 
observed luminosity functions at $3\le z \le 10$. (see also Samui, Subramanian 
\& Srianand (2009) \cite{sam_sub_2009}; hereafter SSS09). 
Here, we use the same approach in order to constrain neutrino masses.
We briefly describe the modeling here and the interested reader 
may refer to SSS07 and SSS09 for more details. 

The star formation rate of an individual dark matter halo of mass $M$
collapsed at redshift $z_c$ and observed at redshift $z$ is modeled by
(see \cite{chi_ost_2000}, \cite{cho_sri_2002})
\begin{eqnarray}
\dot M_{SF}(M,z,z_c) = f_\ast \left(\frac{\Omega_b}{\Omega_m} M \right) 
\frac{t(z)-t(z_c)}{\kappa^2 t^2_{dyn(z_c)}}  \times \exp\left[-\frac{t(z)-t(z_c)}
{\kappa t_{dyn(z_c)}}\right], 
\label{eqn:m_star}
\end{eqnarray}
where $f_\ast$ is the fraction of the total baryonic mass that 
is converted into stars over the entire lifetime of the galaxy. 
Here $t_{dyn}(z_c)$ is the dynamical time scale of a halo collapsing at $z_c$
(Eq.~(3) of SSS07) and $\kappa$ is a parameter which governs the duration of the star 
formation activity in the halo which we take here to be unity. Further, $t(z)$ is 
the age of the universe at redshift $z$; thus $t(z) - t(z_c)$ is the age of the galaxy 
at $z$. We assume that stars are formed with a normal Salpeter IMF in the mass 
range from $1 - 100 ~M_\odot$. The population synthesis code Starburst99 
\cite{starburst_1999} is used to 
obtain the luminosity of a galaxy
undergoing a burst of star formation, as a function of time for
a given rest wavelength of 1500 \AA. The assumed star formation
rate of a galaxy in Eq. (\ref{eqn:m_star}) is then convolved with
this burst luminosity to get the evolution of luminosity ($L_{1500}$)
of an individual star forming galactic halo
(See Eq.~(6) and figure 1 of SSS07).
Only a fraction ($1/\eta$) of the total light produced by the
stars comes out of the galaxy due to the absorption by dust.
We convert this luminosity
($L = L_{1500}/\eta$) to a standard absolute AB
magnitude $M_{AB}$  using the equation given by \cite{oke_gun_1983}.
The luminosity function $\Phi(M_{AB}, z )$ at any redshift $z$ is then given by 
\cite{sam_sri_2007}

\begin{equation}
\Phi(M_{AB}, z ) ~\de M_{AB} = \int\limits_z^\infty \de z_c ~ \frac{\de N_{ST}(M,z_c)}{\de z_c}
\:\:\:\:\:\f{\de M}{\de L_{1500}}~\f{\de L_{1500}}{\de M_{AB}} ~ \de M_{AB}
\label{eqn:lfun}
\end{equation}
where $\de N_{ST}(M,z_c)/\de z_c =  
\dot N_{ST}(M,z_c) dt/dz_c$, and
$\dot N_{ST}(M,z_c) dM$ is the formation rate of objects in the mass
range $(M, M+dM)$ at redshift $z_c$.
We model this formation rate as the time derivative
of ST mass function as (i) they are found to be the best
in reproducing the observed LF of high-$z$ LBGs
(see SSS09),
and (ii) as explained earlier gives good fit to the abundance of
dark matter halos in N-body simulations incorporating massive neutrinos
\cite{bra_han_2010}.

Star formation in a given halo also depends on the cooling efficiency of the gas and 
various feedback processes. We assume that gas in halos with virial temperatures 
($T_{vir}$) in excess of $10^4$ K can cool (due to recombination line cooling from 
hydrogen and helium) and collapse to form stars. However the ionization of the IGM 
by UV photons increases the temperature of the gas thereby increasing the Jean's mass 
for collapse. Thus in ionized regions, we incorporate this feedback by a complete 
suppression of star formation for halos with circular velocity 
$v_c \leqslant 35$ km s$^{-1}$ and no suppression with $v_c  \geqslant V_u = 95$ km s$^{-1}$ 
\cite{bro_loe_2002}. 
For intermediate circular velocities, a linear fit from $1$ to $0$ is 
adopted as the suppression factor (\cite{bro_loe_2002}; 
see also \cite{ben_lac_2002,dij_hai_2004}, SSS07). 
In SSS07 we found that this feedback mechanism naturally leads 
to the observed flattening of the LF at the low luminosity end  
\footnote{In principle, one could also consider star formation in smaller mass halos with 
$T_{\rm vir}\ge300$ K, where the cooling is due to the H$_2$ molecular line emission 
\cite{hai_abe_2000}. While star formation in such halos could be important for reionization 
and IGM metal enrichment, in the post reionization era, it is suppressed by radiative feedback.}. 
Note that implementing radiative feedback in a model requires
a knowledge of the reionization epoch.
A number of observations suggest that 
reionization is nearly complete by a redshift $z_{re} \geq 6$ 
(e.g. \cite{fan_mic_2006, kom_smi_2010}).
We will show below that the strongest constraints on neutrino masses
are obtained from the luminosity function at $z=4$.
As this redshift is much lower than $z_{re}$, our limits
on the neutrino mass then turns
out to be insensitive to the exact $z_{re}$
as long as it is greater than $6$.

In our models, we also incorporate the possible Active Galactic Nuclei (AGN) 
feedback that suppresses star formation 
in the high mass, by multiplying the star formation rate by a factor 
$[1+(M/M_{AGN})^3]^{-1}$, as in SSS07. 
This sharply decreases the star formation activity in 
high mass halos above a characteristic 
mass scale $M_{AGN}$, which is believed to be $\sim 10^{12} M_\odot$ 
(see \cite{bow_ben_2006,bes_kai_2006}). 
The determination of $M_{AGN}$ is done by fitting 
luminosity functions at a fiducial redshift and will 
be discussed below.
In what follows, we compute the high redshift luminosity 
functions of galaxies in the presence of massive 
neutrinos using the above semi-analytical 
prescription and comparing it with observations to 
put limits on neutrino mass. 

\section{Effect of neutrino mass on high redshift LF}
\label{nuLF}

We now illustrate the effect of massive neutrinos on high redshift galaxy UV LF.
We show in Fig.~\ref{fig:lf_zall} our model predictions of luminosity functions 
at different redshifts, $z=3-6$, along with the observational data.
The redshifts are indicated at the top of each panel.
The observed data points are taken from 
Reddy et. al (2007) \cite{red_ste_2007} for $z=3$,
and Bouwens et. al. (2007) \cite{bou_ill_2007} for 
$z = 4,~ 5$ and $6$, and corrected to 
the fiducial WMAP7 cosmology described earlier.
As in SSS09, we 
take into account cosmic variance (\cite{bec_sti_2006}; \cite{bou_ill_2007}) 
and add an uncertainty of 14\% to the Poisson error in quadrature  for 
redshift $4-6$ data. 

The solid line shows the predicted best fit luminosity functions at
various redshifts for the fiducial cosmology and with $\Sigma m_\nu=0$.
In order to fit the observed data points we have adjusted the free parameter
$f_\ast/\eta$ in our models, using $\chi^2$ minimization.
We use all the points 
excluding the last two data points for $z=4,5$ and the last point
for $z=6$, in the low luminosity end 
while fitting the LF, as these points are 
affected by the incompleteness of the survey (see \cite{bou_ill_2007}).
For this illustrative study, we also fix ${\rm V_u}$ = 95 km s$^{-1}$ 
and adopt $M_{AGN}=1.8\times 10^{12}M_\odot$ that best fits the LF at $z=4$ 
with $\Sigma m_\nu = 0$.

Our model with $\Sigma m_\nu = 0$ (solid curve) reproduces 
the observed LF very well at all the redshifts except $z=5$ 
(where SSS09 also earlier pointed out one discrepant
point). The best fit values of $f_\ast/\eta$ at different redshifts 
and the corresponding $\chi^2$ and reduced 
$\chi^2_\nu$, are tabulated in Table~\ref{tab:lum_fit}.
The required values of $f_\ast/\eta$, and reduced $\chi^2_\nu$,
at different $z$ 
are nearly the same and 
similar to what we obtained in
SSS09 (which used slightly different cosmological
parameters). The flattening of the predicted LF as seen 
in Fig.~\ref{fig:lf_zall}
at the faint end is due to the radiative feedback. 

\begin{table}
\begin{center}
\begin{tabular}{|c|   c   c  c|   c   c  c| c c c| }
\hline
&\multicolumn{3}{c|}{$\Sigma m_{\nu} =0$ eV} \T \B & \multicolumn{3}{c|}{$\Sigma m_{\nu} =0.5$ eV} &\multicolumn{3}{c|}{$\Sigma m_{\nu} =1$ eV} \\ \cline{2-10}
{$z$} \HS &$f_\ast/\eta$ \T \B \HS &$\chi^2$  \HS &$\chi^2_\nu$ \HS &$f_\ast/\eta$ \HS &$\chi^2$ \HS &$\chi^2_\nu$ \HS &$f_\ast/\eta$ \HS&$\chi^2$ \HS&$\chi^2_\nu$  \\ \hline 
3.0  \T  &0.040  &15.14  &1.89  &0.052 &19.42  &2.16   &0.086  &31.39   &3.92   \\
4.0  \T  &0.039  &16.09  &1.46  &0.067 &19.49  &1.77   &0.152  &40.08   &3.46   \\
5.0  \T  &0.030  &45.71  &5.07  &0.054 &44.87  &4.99   &0.179  &12.91   &1.43   \\
6.0  \B \T &0.041  &4.34   &0.72  &0.082 &5.64   &0.94   &0.250  &4.97    &0.83  \\
\hline
\end{tabular}
\caption {The best fit values $f_\ast/\eta$, corresponding 
$\chi^2$ and reduced $\chi^2_\nu$ for our models at various 
redshifts for three different values
of $\Sigma m_\nu$. 
}
\label{tab:lum_fit}
\end{center}
\end{table}

{\begin{figure}
\begin{center}
\includegraphics[trim=0cm 0cm 0cm 0cm, clip=true, width =16cm, height=15cm, angle=0]
{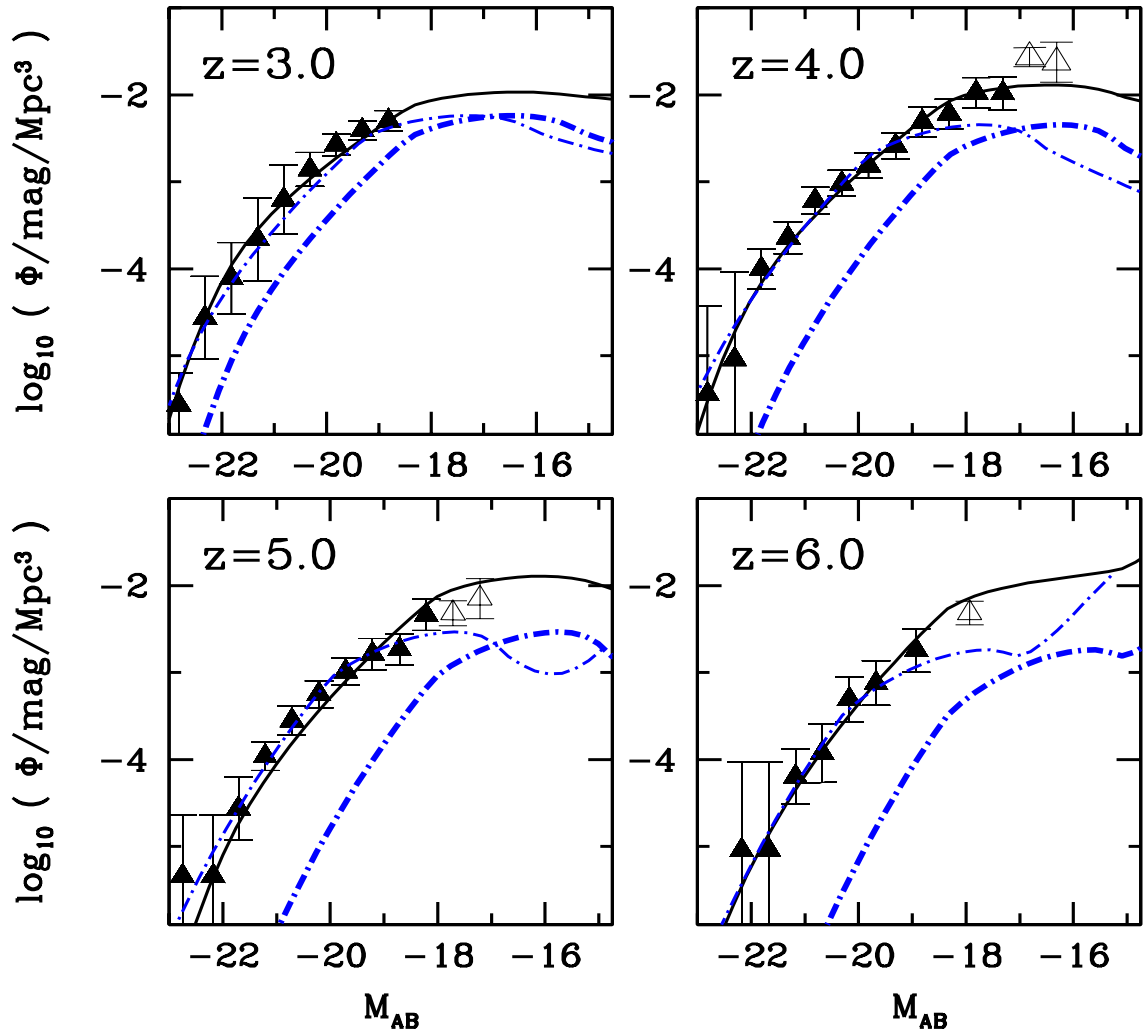} 
\caption{ UV LF of LBGs at redshifts $3,4,5$ and $6$. The solid (black) 
line shows the predicted best fit LF 
for our model with $\Sigma m_\nu=0$.  The thick dashed-dotted (blue) 
curve shows our model predictions with $\Sigma m_\nu = 1$ eV 
and using the same best fit 
$f_*/\eta$ as the $\Sigma m_\nu=0$ eV model. 
This is to illustrate the suppression due to 
massive neutrinos. 
The thin dashed-dotted (blue) 
curves are best fits for the models 
with $\Sigma m_\nu = 1$ eV. 
The data points (filled and open triangles) for $z=3$ are taken from 
\cite{red_ste_2007} and for $z=4-6$ are from \cite{bou_ill_2007}. 
}
\label{fig:lf_zall}
\end{center}
\end{figure}

The thick dashed-dotted (blue) curves show the 
predicted LF if we use the same $f_*/\eta$ for a MDM model 
with $\Sigma m_\nu = 1.0$ eV. 
It is clear that there is an order of magnitude suppression 
in the number density of galaxies at 
a given luminosity, 
which increases with increasing redshift. 
This is because the presence of neutrinos
suppresses the formation rate of halos
at the mass and redshift scales
of our interest. 
We can 
make our model predictions match with the observed data  
by increasing $f_*/\eta$ (i.e shifting this curve along the luminosity axis). 
These best fit values of $f_\ast/\eta$ at different redshifts 
for the case $\Sigma m_\nu=1$ eV, and the corresponding reduced 
$\chi^2_\nu$, are also tabulated in Table~\ref{tab:lum_fit}.
Firstly, we can see from Table~\ref{tab:lum_fit}, 
that at each z the value of $f_\ast/\eta$ (or light to mass ratio) 
needed to fit the observed luminosity function increases with $\Sigma m_\nu$.  
For example, at $z=6$ one needs $\sim 6$ times more baryons
to convert into stars for a model with $\Sigma m_\nu=1$ eV compared
to the zero neutrino mass case. 
Furthermore, for $\Sigma m_\nu=1$ eV, 
the value of $f_\ast/\eta$ required to fit the UV LFs also
increases significantly with $z$, in contrast to the zero neutrino mass case.
Therefore, any independent constraint on $f_*/\eta$, especially
at high $z$, could lead to useful constraints on the neutrino mass.

More importantly, these best fit luminosity functions, 
obtained with the new $f_*/\eta$ 
(thin dashed-dotted curves), have a very different shape
compared to the zero neutrino mass case. 
In particular, the predicted luminosity
functions in models with $\Sigma m_\nu=1$ eV, are suppressed
at the low luminosity end compared to the zero mass case. 
This is basically because increasing $f_\ast/\eta$ increases
the light to mass ratio, which brings even small mass galaxies
whose star formation has been suppressed due to radiative feedback,
into the observable luminosity range. Therefore strong constraints on the
neutrino mass can in principle be obtained by comparing the shape of
the predicted luminosity functions with observations,
independent of the free parameter $f_*/\eta$, provided
the observations cover a wide luminosity range and in
particular extend to the very faint end of the LF.

This requirement is at present best satisfied at $z=4$. 
This is because, the observed LF for $z=4$ is well defined over a wide range 
of luminosity thanks to the Hubble Ultra 
deep field (HUDF) and the Great Observatories Origins Deep Survey (GOODS) 
(\cite{bec_sti_2006,bou_ill_2007}). This redshift 
is also well below the redshift of reionization 
\cite{fan_mic_2006,kom_smi_2010}; 
so radiative feedback can be applied without any ambiguity. 
Therefore for demonstrating that quantitative limits on $\Sigma m_\nu$ can 
be obtained with well defined LF, we
will concentrate below on the LF of LBGs at $z=4$.

At this redshift we already see from Table~\ref{tab:lum_fit} that, 
while $\chi^2=16.09$ for $\Sigma m_\nu=0$ eV model, it increases 
by $\Delta\chi^2 \sim 3.4$ for $\Sigma m_\nu=0.5$ eV
and by $\Delta\chi^2 \sim 24$ for $\Sigma m_\nu=1$ eV models.
This suggests that $\Sigma m_\nu \sim 1$ eV is strongly disfavored
at a 5$\sigma$ level, and a typical 
$2\sigma$ upper limit is close to $\Sigma m_\nu \sim 0.5$ eV.
While this is encouraging, we have been 
using a fixed set of cosmological parameters
and therefore one needs to check if such a conclusion also follows when we
vary these parameters. We examine this issue further below.

\section{Limits on neutrino mass}
\label{result}

In the previous section we adopted a fiducial cosmology to
examine the effect of a non-zero neutrino mass on the LF of LBGs.
In order to obtain quantitative upper limits by exploring 
the full range of cosmological 
parameters consistent with both the WMAP7 data and the 
observed LF of LBGs, we have
performed a Markov Chain Monte Carlo (MCMC) analysis using 
the publically available CosmoMC 
code \cite{cosmomc_2002} (we use the CosmoMC May 2010 version 
and the WMAP likelyhood code version 4.1).
The default version of CosmoMC 
constrains the neutrino 
fraction in dark matter defined as $\bar{f}_\nu = \Omega_\nu/\Omega_{DM}$ 
where $\Omega_{DM} = \Omega_{CDM}+\Omega_\nu$.
So we constrain $\bar f_\nu$ and obtain neutrino mass 
as $\Sigma m_\nu = 93.14\bar f_\nu \Omega_{DM} h^2$, 
where we use the mean value of $\Omega_{DM}$ from our MCMC analysis. 
We will also carry out the MCMC analysis by giving a prior
in terms of $\Sigma m_\nu$ to directly constrain its value.
In addition to $\bar f_\nu$ (or $\Sigma m_\nu$), we explored the usual 7 
dimensional parameter space 
($\Omega_b h^2, \Omega_c h^2, \theta_s,\tau,n_s,A_s,A_{SZ}$).

In order to combine the constraints from the UV LF data with
other constraints (like the WMAP7 data),
we have added a new module to CosmoMC. This module
computes the likelihood for
LF of LBGs by comparing the theoretically predicted LF for any set of 
cosmological parameters with observed LF. 
The observed data points at $z=4$ by \cite{bou_ill_2007}, 
are for a cosmology with $\Omega_m =0.3$ and $\Omega_\Lambda=0.7$ and $h=0.7$.
Therefore for each set of cosmological parameters in the chain, 
we first correct the observed data to that cosmology, by modifying the
volume and distance scales appropriately. Then for each member of the chain
we compute the theoretical LF, compare it with the observed data,
and minimize the $\chi^2$ by varying $f_*/\eta$. This minimum $\chi^2$ is
used for calculating the LF likelihood.

\begin{table}
%\scriptsize
\small
\begin{tabular}{l l l l l l l l l l}
\hline  
\hline  
Parameter \HSN&WMAP7 \T\HS&WMAP7  \HS&WMAP7 \HS&WMAP7   &WMAP7      &WMAP7     &WMAP7    &WMAP7  \HS&WMAP7     \\
            &           &+LF       &+HST     &+LF     &($m_\nu$)  &+LF       &+LF      &+LF(FB)   &LF(FB)       \\
            &         \B&          &         &+HST    &           &($m_\nu$) &+HST     &          &+HST      \\
            &           &          &         &        &           &          &($m_\nu$)&          & \\
\hline
\hline
$10^2\Omega_bh^2$     \T&$2.223^{+0.060}_{-0.058}$  &$2.229^{+0.055}_{-0.056}$   &$2.263^{+0.050}_{-0.056}$    &$2.259^{+0.054}_{-0.054}$  &$2.245^{+0.060}_{-0.059}$ 
                              &$2.229^{+0.055}_{-0.056}$  &$2.257^{+0.053}_{-0.053}$   &$2.235^{+0.056}_{-0.055}$    &$2.266^{+0.051}_{-0.052}$    \\[1.0ex]

$10\Omega_{DM}h^2$    &$1.172^{+0.070}_{-0.068}$  &$1.228^{+0.065}_{-0.064}$   &$1.107^{+0.051}_{-0.050}$    &$1.169^{+0.056}_{-0.058}$  &$1.177^{+0.071}_{-0.069}$
                              &$1.235^{+0.065}_{-0.064}$  &$1.168^{+0.056}_{-0.058}$   &$1.208^{+0.060}_{-0.059}$    &$1.156^{+0.045}_{-0.045}$    \\[1.0ex]

$\tau$                        &$0.087^{+0.014}_{-0.014}$  &$0.090^{+0.013}_{-0.014}$   &$0.091^{+0.015}_{-0.015}$    &$0.095^{+0.015}_{-0.014}$  &$0.086^{+0.014}_{-0.014}$
                              &$0.090^{+0.013}_{-0.014}$  &$0.095^{+0.015}_{-0.014}$   &$0.090^{+0.013}_{-0.014}$    &$0.093^{+0.014}_{-0.015}$  \\[1.0ex]

$n_s$		              &$0.962^{+0.016}_{-0.015}$  &$0.961^{+0.015}_{-0.014}$   &$0.973^{+0.013}_{-0.013}$    &$0.969^{+0.014}_{-0.016}$  &$0.962^{+0.015}_{-0.015}$
                              &$0.961^{+0.015}_{-0.014}$  &$0.970^{+0.014}_{-0.016}$   &$0.963^{+0.014}_{-0.014}$    &$0.971^{+0.013}_{-0.012}$  \\[1.0ex]

$\sigma_8$                    &$0.717^{+0.071}_{-0.072}$  &$0.820^{+0.051}_{-0.052}$   &$0.756^{+0.049}_{-0.047}$    &$0.821^{+0.042}_{-0.041}$  &$0.717^{+0.071}_{-0.072}$ 
                              &$0.824^{+0.051}_{-0.050}$  &$0.821^{+0.042}_{-0.043}$   &$0.808^{+0.038}_{-0.038}$    &$0.813^{+0.030}_{-0.029}$   \\[1.0ex]

$H_0$                         &$66.1^{+4.1}_{-4.9}$       &$65.5^{+2.8}_{-2.8}$        &$70.3^{+2.5}_{-2.5}$         &$68.7^{+2.1}_{-2.2}$       &$66.0^{+4.2}_{-4.0}$
                              &$65.3^{+2.9}_{-2.8}$       &$68.7^{+2.2}_{-2.2}$        &$66.3^{+2.9}_{-2.9}$         &$69.2^{+2.2}_{-2.2}$    \\[1.0ex]

$f_\ast/\eta$       &\HS---                     &$0.035^{+0.007}_{-0.007}$   &\HS---                       &$0.036^{+0.007}_{-0.007}$        &\HS---
                              &$0.034^{+0.005}_{-0.006}$        &$0.035^{+0.007}_{-0.006}$                &$0.037^{+0.005}_{-0.005}$    &$0.037^{+0.005}_{-0.005}$ \\[1.0ex]

$\bar{f}_\nu$                 &$<0.091$                   &$<0.048$                    &$<0.052$                     &$<0.028$                   &\HS--- 
                              &\HS---                     &\HS---                      &$<0.042$                     &$<0.026$ \\[2.8ex]

$\Sigma m_\nu$(eV)            &$<1.0$                 \B &$<0.55$                     &$<0.54$                      &$<0.31$                    &$<1.08$
                              &$<0.52$			  &$<0.29$ 		       &$<0.48$  		     &$<0.28$ \\
\hline
\hline

\end{tabular}
\caption{
Results of our MCMC analysis to constrain $\Sigma m_\nu $.  
The first column lists the set of parameters that are obtained 
from our MCMC analysis. The last two rows give limits on 
$\bar{f}_\nu$ or $\Sigma m_\nu$, at the $95\%$ CL.  
For cases where we constrain $\bar{f}_\nu$, the neutrino mass is given by 
$\Sigma m_\nu = 93.14 \bar{f}_\nu \Omega_{DM} h^2$,
where for $\Omega_{DM} h^2$ we use the mean value given in row 3.
For other parameters their mean values and 1$\sigma$ range are given.
}
\label{table:cosmomc_result}
\end{table}

\begin{figure*}
\subfigure[Best fit Cls]{
\includegraphics[trim=0cm 0cm 0.0cm 0.0cm, clip=true, width =8.5cm, height=6.5cm, angle=0]
{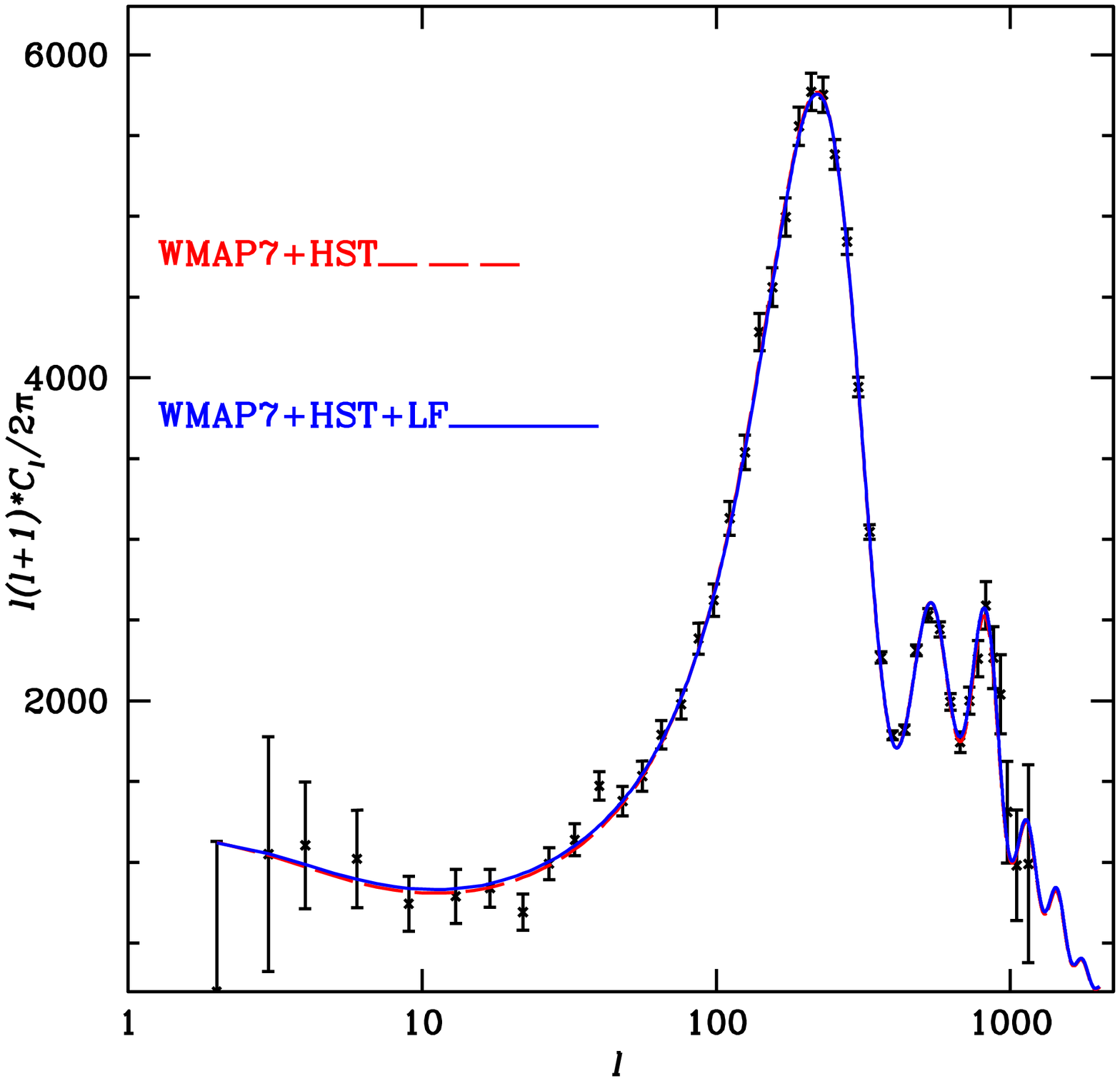} \label{fig:cls_bf}}
\subfigure[Best fit LF at z=4]{
\includegraphics[trim=0cm 0cm 0.0cm 0.0cm, clip=true, width =8.5cm, height=6.5cm, angle=0]
{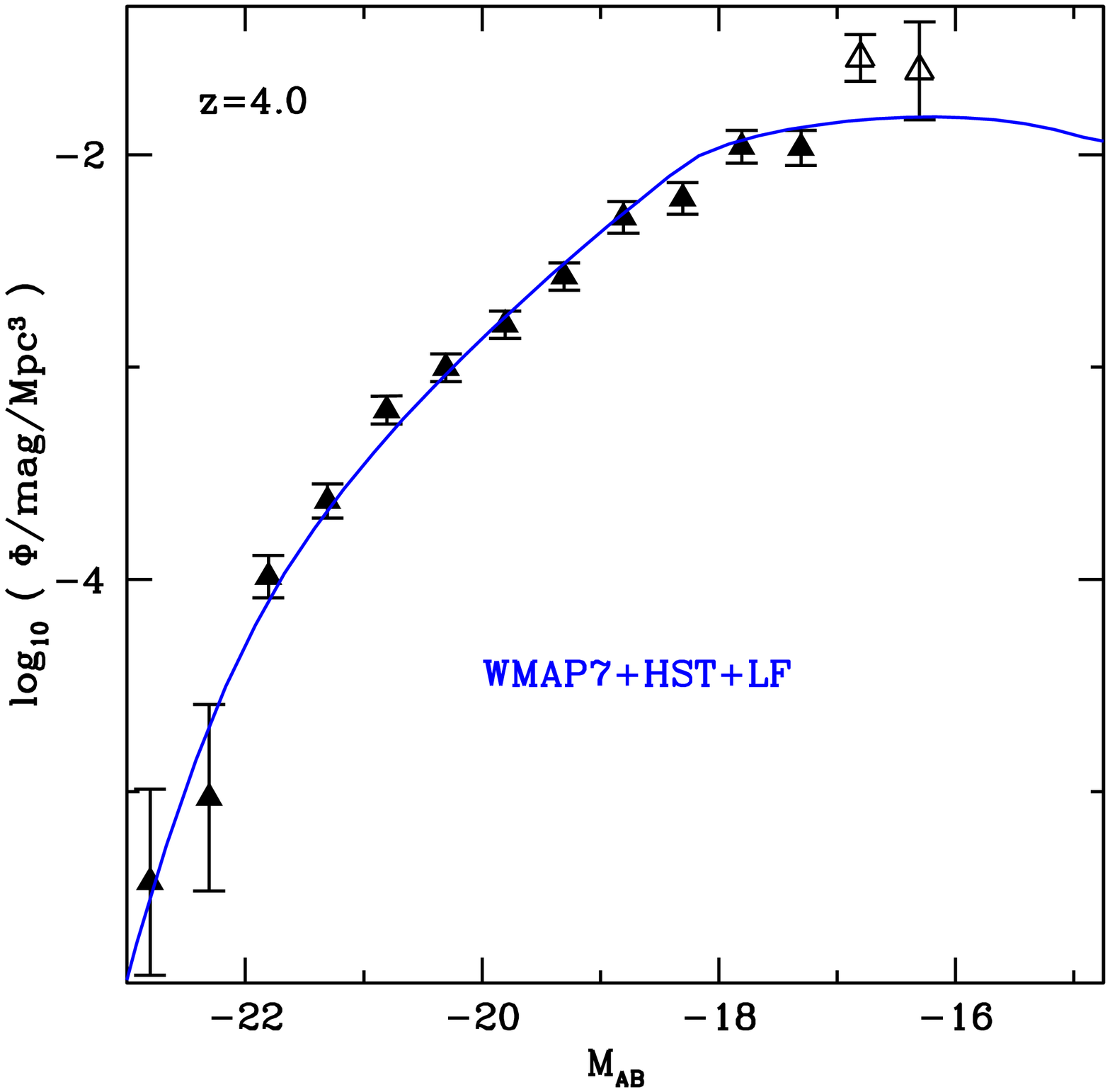} \label{fig:lfun_bf}}
\caption{The fits to the CMBR TT power spectrum (Cls) (left panel) and the 
LF at $z=4$ (right panel) using the best fitted values of 
cosmological parameters obtained from
our MCMC analysis. 
Solid blue lines are obtained for cosmological 
parameters corresponding to WMAP7+HST+LF data. 
The red dashed line in the left panel gives for comparison, 
Cls for best fitted cosmological parameters based on only 
WMAP7+HST data. 
}
\label{fig:best_fit}
\end{figure*}

\begin{figure}
\begin{center}
\includegraphics[trim=0cm 0cm 0cm 0cm, clip=true, width =15cm, height=12cm, angle=0]
{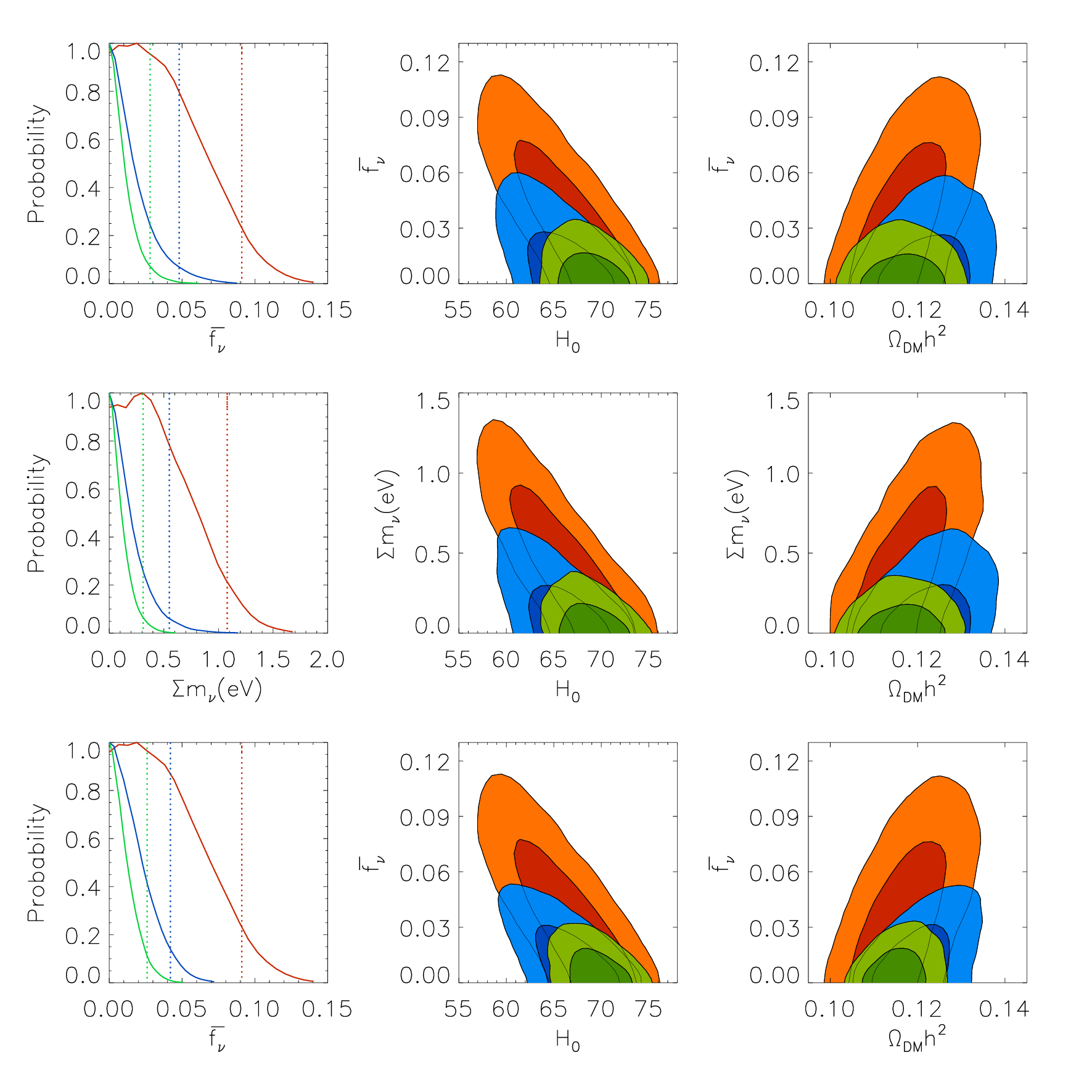} 
\caption{ Various 1D and 2D marginalized distributions from our 
analysis. The top panel  
corresponds to our fiducial LF model with feedback parameters 
$M_{AGN}=1.8 \times 10^{12} M_\odot$ 
and $V_u=95$ km s$^{-1}$. The middle panel shows plots corresponding 
to constraining $m_\nu$ directly with
same feedback parameters as above. The bottom panel is for 
the  LF model with feedback parameters 
$M_{AGN}=1.5 \times 10^{12} M_\odot$ and $V_u=105$ km s$^{-1}$ (model FB). 
In each row the left panel gives the marginalized 1D 
distribution for $\bar{f}_\nu$ or $m_\nu$. 
The vertical dashed lines correspond to 95\% confidence levels.
The other two panels in each row, 
show the regions of 68\% (dark color) and 95\%
(light color) confidence levels 
for $H_0$ and $\Omega_{DM}h^2$ against $\bar{f}_\nu$ or $m_\nu$. 
The various contours corresponds to constraints obtained using
WMAP7-only (red), WMAP+LF (blue) and WMAP+HST+LF (green) data. 
}
\label{fig:banana}
\end{center}
\end{figure}

We give the constraints on $\Sigma m_\nu$ obtained from
the MCMC analysis in Table~\ref{table:cosmomc_result} and 
Fig~\ref{fig:banana}. The convergence of MCMC chains were
diagnosed using the usual Gelman-Rubin statistic and an $R-1 < 0.01$
was achieved in general. 
First we have examined the constraints from
WMAP7 data alone \cite{kom_smi_2010,jar_ben_2010}. 
In this case we obtain an upper limit $\bar{f}_\nu < 0.091$
at the 95\% CL, which when converted to a mass limit
as described above, gives $\Sigma m_\nu < 1.0$ eV. The other cosmological
parameters, given in column 2 of  Table~\ref{table:cosmomc_result},
are almost identical to that obtained by 
\cite{arch10} for their case of sudden reionization.
Note that constraining $\Sigma m_\nu$ directly gives 
a very similar limit $\Sigma m_\nu < 1.08$ eV at 95\% CL with 
almost the same values for the other cosmological parameters 
(see column 6 of Table~\ref{table:cosmomc_result}).

Combining the UV luminosity function data at $z=4$ with the 
WMAP7 data gives a significantly lower limit on $\Sigma m_\nu$.
We get $\bar{f}_\nu < 0.048$ corresponding to $\Sigma m_\nu \leq 0.55$ eV
at the  95\% CL. 
Directly constraining the neutrino mass also leads to 
a very similar limit, $\Sigma m_\nu < 0.52$ eV at the 95\% CL.
Thus the limit on the neutrino mass is decreased
by a factor of $\sim 2$ by the addition of the constraints from
the $z=4$ UV LF. The cosmological
parameters for these two cases, given respectively 
in column 3 and 7 of  Table~\ref{table:cosmomc_result},
are very similar. These parameters are also similar
to the parameters obtained from WMAP7 alone (column 2 
of  Table~\ref{table:cosmomc_result}),
except for $\sigma_8$, which is increased on including the
constraints from the LF data.
Note that these limits on $\Sigma m_\nu$, obtained by varying
all the cosmological parameters, bear out the naive
expectation from the analysis of the Section~\ref{nuLF}, where we
fixed the cosmology.

Next we examined the effect of adding the constraints from the 
$H_0$ determination of the HST SHOES (Supernova H0 for the Equation of State) 
program (which we refer to as HST) 
(\cite{rie_mac_2009}; see also \cite{fre_mad_2001}).
First WMAP7+HST data alone leads to a tighter neutrino mass
limit of $\bar{f}_\nu < 0.052$ (or $\Sigma m_\nu < 0.54$) at 95\% CL,
consistent with the earlier results of \cite{sek_ich_2010}.
On adding in the constraint from UV luminosity function data at $z=4$
this limit is also further decreased. We obtain  
$\bar{f}_\nu < 0.028$ corresponding to $\Sigma m_\nu \leq 0.31$ eV.
(The corresponding cosmological parameters are given 
in column 4 and 5 of  Table~\ref{table:cosmomc_result} respectively). 
Again directly constraining the neutrino mass gives 
a very similar limit, $\Sigma m_\nu < 0.29$ eV at the 95\% CL.
Therefore, we see that adding in the LF data further 
decreases the limit on the neutrino mass by another factor of $\sim 2$.
This limit, $\Sigma m_\nu < 0.29$ eV is almost a factor $\sim 4$
improvement compared to the limit obtained using the WMAP7 data alone.
The other cosmological parameters for this case are given in
column 8 of Table~\ref{table:cosmomc_result} and 
Fig~\ref{fig:best_fit} shows the corresponding 
predicted TT power spectrum of CMBR (Cls) 
and the LF at $z=4$ overplotted against the data.
In Fig~\ref{fig:best_fit} we also plot the Cls corresponding to
the parameters which do not include the LF constraint 
(column 4 of Table~\ref{table:cosmomc_result}), which shows that 
Cls for these two cases are almost identical.

All the above neutrino mass limits along with the other 
cosmological parameters are summarized
in Table~\ref{table:cosmomc_result}.
The corresponding 1D marginalized 
distribution for $\bar f_\nu$, from various MCMC analysis, 
is shown in the top left panel of Fig.~\ref{fig:banana}. 
The top right 2 panels of Fig.~\ref{fig:banana} shows 68\% and 95\% 
marginalized distributions for $H_0$ and $\Omega_{DM}h^2$ against 
$\bar{f}_\nu$, the massive neutrino fraction of the dark matter.
The results of directly constraining $\Sigma m_\nu$ is shown
in the middle panels of Fig.~\ref{fig:banana}.  
We see also from these figures adding the
constraint from the $z=4$ UV luminosity function significantly
improves the constraint on neutrino masses.

In the above analysis we fixed the feedback parameters to
$M_{AGN}=1.8 \times 10^{12} M_\odot$ and $V_u=95$ km s$^{-1}$,
as described earlier. It is of interest to test the sensitivity
of our neutrino mass limits to changes in these values.
One way of doing this would be to vary 
$M_{AGN}$, $V_u$ and $f_*/\eta$ as free parameters 
to obtain the best fit LF and its corresponding likelihood 
for each step of the MCMC analysis. However, 
implementing the above 
procedure in CosmoMC (where the cosmology is also varied) 
is computationally expensive, and is outside the scope of
the present work.
Note however that the feedback parameters are expected 
to depend on physical nature of the feedback mechanisms and not
on cosmology. 
Therefore, as an alternative to varying the feedback parameters
within CosmoMC, we adopt our fiducial cosmology and 
carry out the following procedure.
We vary $M_{AGN}$, $V_u$ and $f_*/\eta$ in the parameter ranges 
$10^{11} M_\odot \le M_{AGN} \le 5\times 10^{12} M_\odot $, 
$85 \ {\rm km s}^{-1} \le V_u \le 120 \ {\rm km s}^{-1}$ 
and $0.005 \le f_*/\eta \le 1$, 
to obtain the best fit LF and its corresponding $\chi^2$ 
as a function of $\Sigma m_\nu$. This best fit $\chi^2$ as a function
of $\Sigma m_\nu$,
obtained after varying all 3 free parameters given above, is
shown in Fig. ~\ref{fig:lf_chi2}. It is clear from Fig. ~\ref{fig:lf_chi2}
that even after varying the feedback parameters,
the global minimum value of $\chi^2$ obtains for the 
zero neutrino mass case. This global minimum obtains for
$V_u = 105$ km s$^{-1}$ and $M_{AGN}= 1.5\times 10^{12}M_\odot$.
The typical 2$\sigma$ limit is $\Sigma m_\nu = 0.35$ eV, 
even in this case where all the feedback parameters are varied. 
This excercise suggests that the uncertainity in the feedback parameters
may not have a strong influence on the the upper limit on $\Sigma m_\nu$.

\begin{figure}
\begin{center}
\includegraphics[trim=0cm 0cm 0cm 0cm, clip=true, width =7cm, height=6cm, angle=0]
{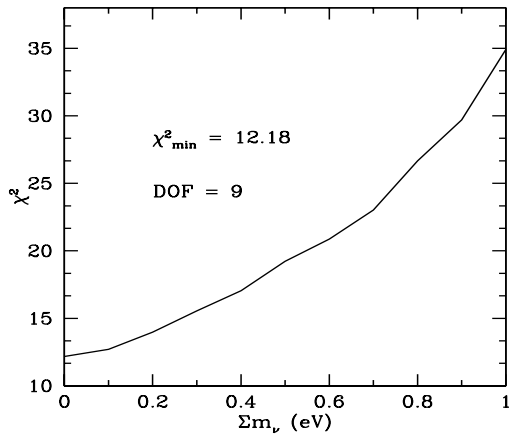} 
\caption{ The $\Sigma m_\nu$ vs $\chi^2$ curve for our fiducial cosmology with massive neutrinos. 
Here we varied $M_{AGN},~V_u$ and $f_\ast/\eta$ as free parameters in our model for
LF. The solid black curve gives best fit $\chi^2$ as a function of $m_\nu$ after marginalizing over
all the above three parameters. The number of degrees of freedom is 9. 
}
\label{fig:lf_chi2}
\end{center}
\end{figure}

We have repeated our MCMC analysis adopting these best fit 
values of $M_{AGN}$ and $V_u$ (which we label as model FB).
The results of the analysis, adopting the feedback parameters of model FB, 
are presented in the last 2 columns 
of Table~\ref{table:cosmomc_result} and in the bottom panels of
Fig.~\ref{fig:banana}. Combining the WMAP7 data and the LF data now
leads to a lower limit on the neutrino fraction $\bar{f}_\nu < 0.042$ 
and a corresponding $\Sigma m_\nu \leq 0.48$ eV. 
On adding also the constraints from the 
$H_0$ determination of the HST SHOES program (together with
WMAP7 and UV LF), we get  
$\bar{f}_\nu < 0.026$ corresponding to $\Sigma m_\nu \leq 0.28$ eV
at the $95\%$ CL. We note that these
limits on $\Sigma m_\nu$ are almost identical to that obtained 
using the fiducial feedback parameters. 

In passing we note that we can also get quantitative measurements 
of the astrophysical parameter $f_\ast/\eta$ at $z=4$ from our MCMC analysis.
These values are given in Table~\ref{table:cosmomc_result}, while in
Fig ~\ref{fig:fstar_eta} we plot the 1D marginalized 
probability distributions (PDF) of $f_\ast/\eta$ obtained from the MCMC analysis
for a number of models.  
The red curves are for the fiducial feedback
parameters ($M_{AGN}=1.8 \times 10^{12} M_\odot$ and $V_u=95$ km s$^{-1}$)
and blue curves are for the model FB 
($M_{AGN}=1.5 \times 10^{12} M_\odot$ and $V_u=105$ km s$^{-1}$). 
In particular the solid (red) and
dashed (blue) curves give the PDF 
of $f_\ast/\eta$ when WMAP7 and LF data are
combined. The dotted (red) and dashed-dotted (blue) curves are obtained when 
constraints from WMAP7, HST prior on $H_0$ and LF are combined. 
From this figure and Table~\ref{table:cosmomc_result}, 
we see that $f_*/\eta$ is constrained in a narrow range of
values, with a mean $f_*/\eta \sim 0.034-0.037$. 
If one adopts a value
of $\eta \sim 4$ at $z=4$ \cite{reddy}, then this corresponds to an 
$f_* \sim 0.14-0.15$. Thus about $15\%$ of baryons need
to be converted to stars over several
dynamical timescales, to explain the observed UV LF at $z=4$.
This conclusion obtains regardless
of the feedback parameters and although we have now explored
the full range of cosmological parameters consistent with
CMBR, HST and the LF data. 

\begin{figure}
\begin{center}
\includegraphics[trim=0cm 0cm 0cm 0cm, clip=true, width =8cm, height=8cm, angle=0]
{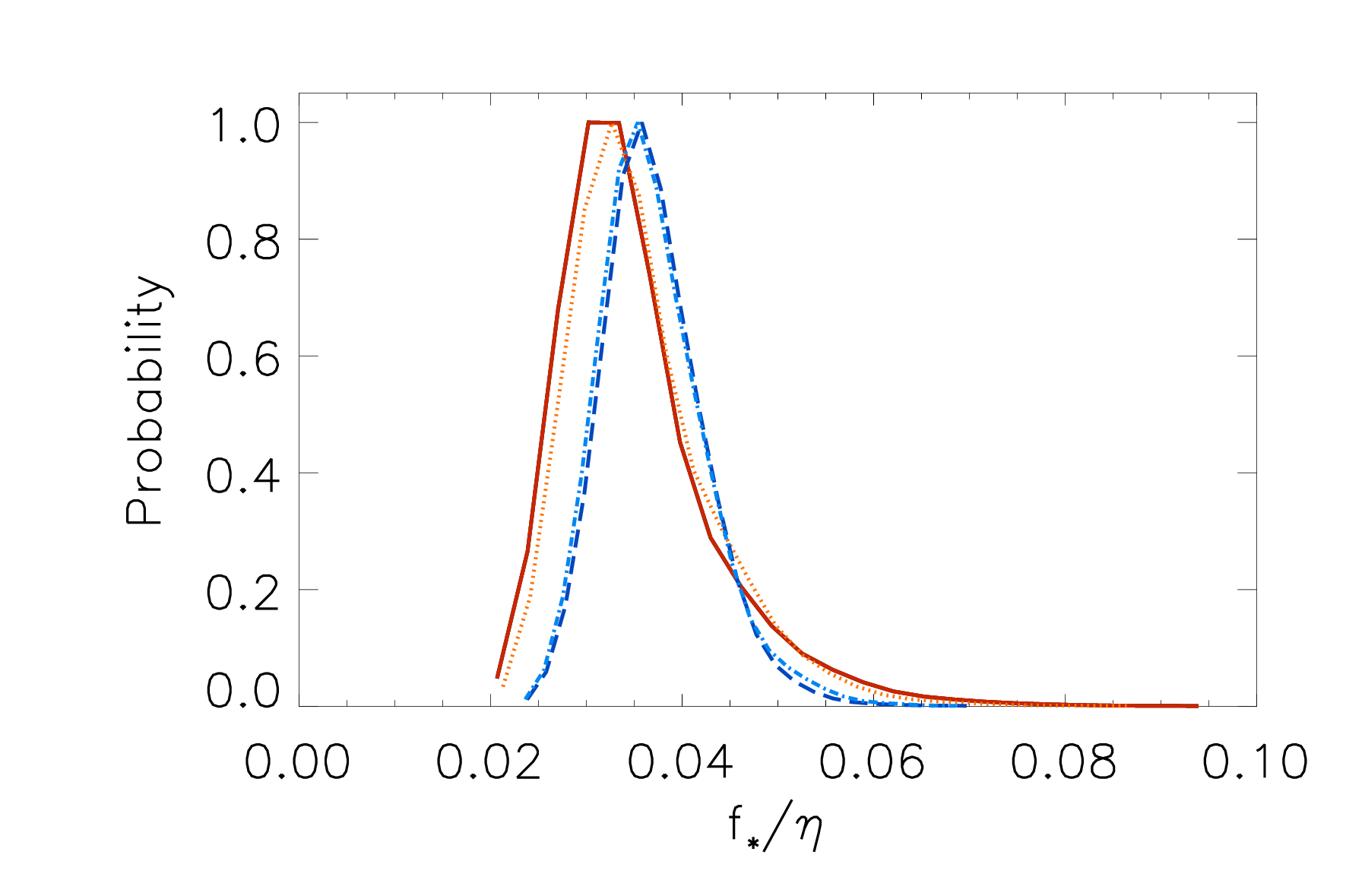} 
\caption{The 1D marginalized distributions of $f_\ast/\eta$ from our analysis using LF at $z=4$. 
The solid and dotted (red) curves respectively gives the PDF
of $f_\ast/\eta$ when one uses 
WMAP7+LF and WMAP7+HST+LF data with the feedback 
parameters as in the fiducial LF model.
The blue dashed and dashed-dotted curves 
give the PDF of $f_\ast/\eta$ 
when WMAP7+LF and WMAP7+HST+LF data are used, and 
for the  LF model with feedback parameters FB. 
}
\label{fig:fstar_eta}
\end{center}
\end{figure}

The results of our MCMC analysis in this section 
shows therefore that fitting
the UV LF of high $z$ galaxies can significantly
strengthen constraints on the neutrino mass.

\section{Conclusions}
\label{conclu}

We have proposed here a novel probe of neutrino masses which 
uses the high redshift UV LF of Lyman break galaxies. 
In particular, our models  constructed in the framework of MDM cosmology, 
show that the observed shape of the UV LF of high redshift LBGs 
can be used to constrain the mass of the 
neutrinos ($\Sigma m_\nu$). 

In the presence of massive neutrinos, the matter power
spectrum is suppressed below the free-streaming scale. 
This in turn results in a suppression
of the abundance of collapsed halos capable of hosting galaxies 
at high redshift, compared to a model where $\Sigma m_\nu=0$. 
We have studied how this affects the UV LF of LBGs from 
$z=3-6$. The UV luminosity of a galaxy in our models depends on the
parameter $f_*/\eta$, where $f_*$ is the total fraction of the baryons
converted into stars and $1/\eta$ is the fraction of the total
light which comes out of the galaxy after taking into account 
dust extinction. For a given $f_*/\eta$, 
we find that the number
density of galaxies at a given luminosity is suppressed 
for models with a non-zero neutrino mass compared
to the zero neutrino mass case. For example,
if $\Sigma m_\nu=1$ eV this suppression factor is about
an order of magnitude at all redshifts, and increases
with redshift. In order to best fit the observed UV LF for
the models with non zero neutrino mass, 
the light to mass ratio of each galaxy, governed by
the parameter $f_*/\eta$ has to be correspondingly 
larger at any redshift. 
At $z=6$ for example, we find that $f_*/\eta$ has to
be 6 times larger for a model with $\Sigma m_\nu=1$ eV compare
to the zero neutrino mass case. 
Moreover, unlike the $\Sigma m_\nu=0$ case, 
the best fit $f_\ast/\eta$ also increases with $z$,
for models with a non zero $\Sigma m_\nu$.
Thus independent constraints on $f_*/\eta$, especially
at high $z$, could lead to useful constraints on the neutrino mass.

%%%&

More importantly, these best fit luminosity functions
with a non-zero $\Sigma m_\nu$, obtained by increasing the light to mass 
ratio of each galaxy (or $f_*/\eta$) brings even small 
mass galaxies, which suffer from
various feedback suppression effects, into the observable range.
We have shown here that this results in a sensitivity of the 
LF shape to the neutrino mass.
The well measured UV LF of LBGs at $z \sim 4$ is best suited
for obtaining quantitative constraints as the data extends
over the widest range of luminosities and 
to much fainter levels than at other redshifts. 
To obtain quantitative upper limits on $\Sigma m_\nu$,
that also explores the full range of cosmological 
parameters, we have carried out an MCMC analysis using 
the publically available CosmoMC 
code \cite{cosmomc_2002}, after adding also a module
which calculates the LF likelihood.
The neutrino mass limits obtained here and other cosmological
parameters for these models are summarized in 
Table~\ref{table:cosmomc_result}.
The corresponding 1D and 2D marginalized 
distribution for $\bar f_\nu$ (or $\Sigma m_\nu$), from various MCMC analysis, 
are shown in Fig.~\ref{fig:banana}. 
 
First, combining the constraints from the WMAP7 data and the 
UV LF of LBGs at $z \sim 4$, we obtain 
a constraint on the neutrino fraction $\bar{f}_\nu < 0.048$ and a
corresponding limit on sum of neutrino masses, $\Sigma m_\nu < 0.55$ eV 
at the 95 \% CL. 
Directly constraining $\Sigma m_\nu$
in the MCMC analysis leads a similar upper limit $\Sigma m_\nu < 0.52$ eV 
at the 95 \% CL. 
Thus the neutrino mass limit is not greatly
sensitive to whether one specifies a prior on $\bar{f}_\nu$ or 
$\Sigma m_\nu$.
We have also tested the sensitivity of these limits to the 
feedback parameters. Adopting a different set of feedback parameters (model FB) 
we get $\bar{f}_\nu < 0.042$ or $\Sigma m_\nu \le 0.48$ eV at the 95 \% CL.
Thus the uncertainties related to halo mass (or circular velocity)
range over which suppression of star formation due to feedback takes place 
and exact value of $M_{AGN}$, do not introduce significant uncertainty in the 
upper limit on $\Sigma m_\nu$. 
Our constraints on $\Sigma m_\nu$, obtained
by combining the $z=4$ LF and the WMAP7 data, are a factor of 2 more stringent
than constraints that would obtain by using the WMAP7 data by itself.

We have also examined the effect of adding the constraints from the
HST prior on $H_0$. Consistent with the earlier work of \cite{sek_ich_2010},
we find that WMAP7+HST data alone leads to a tighter neutrino mass
limit of $\bar{f}_\nu < 0.052$ (or $\Sigma m_\nu < 0.54$) at 95\% CL.
This limit is also further decreased to
$\bar{f}_\nu < 0.028$ or $\Sigma m_\nu < 0.31$ eV
on adding in the constraints from the $z=4$ LF data.
Directly constraining the neutrino mass gives 
in this case $\Sigma m_\nu < 0.29$ eV at the 95\% CL.
Adopting model FB for the feedback parameters also gives
a similar limit  
$\bar{f}_\nu < 0.026$ or
$\Sigma m_\nu < 0.28$ eV at the 95 \% CL.
We note that these neutrino mass limits are almost a factor $\sim 4$
improvement compared to the limit obtained using the WMAP7 data alone.

A summary of the current cosmological and astrophysical constraints
on neutrino mass can be found in \cite{abaz_2011}.
Some of the specific results on $\Sigma m_\nu$ limits are given by 
\cite{tho_abd_2010,rei_ver_2010,seljak_2006,veil_2010,man_all_2010,swa_per_2010,
ter_sch_2008,li_liu_2009,ich_tak_2009}.
The constraints on $\Sigma m_\nu$ obtained here adding in the 
$z\sim4$ UV LF data to the WMAP7 and HST data, are comparable
(or better in several cases) to the above limits.
Our work is mainly a demonstrative first step,
where we have suggested the utility of 
the LF of high redshift galaxies to constrain $\Sigma m_\nu$.
We have concentrated here on the $z=4$ UV LF as this is well
defined over the widest range of luminosities.
%and $z=4$ is well below the redshift of reionization.
Improvements in the LF data, especially at the faint end,
and at higher redshifts together with a better understanding
of the astrophysics of galaxy formation, will allow
us to place more stringent constraints on $\Sigma m_\nu$.

\section*{Acknowledgments}
CJ thanks CSIR for providing support for this work.
We thank Jacob Brandbyge and Toyokazu Sekiguchi for helpful
correspondence. We also thank Moumita Aich, Minu Joy, Toyokazu Sekiguchi and 
Tarun Souradeep for helpful advice on CosmoMC.

%###################################################################################
%%#############################################################################################

\end{document}